# On Slow Dynamic Elasticity at short times

SangMin Lee[1] and Richard L Weaver[2]

[1]Department of Civil and Environmental Engineering, University of Illinois, Urbana, IL 61801
[2]Department of Physics, University of Illinois, Urbana, IL 61801

**Abstract**   It has been reported that slow dynamic nonlinear elastic relaxations, widely thought to proceed in proportion to the logarithm of time since mechanical conditioning ceases, recover at a diminished rate at early times, with a time of transition that varies with the grain size of the material. Here we recount new observations at short times, in the single bead system, in cement paste and in sandstone and mortar.  Notwithstanding the limits imposed by finite duration ring down such that the effective instant of conditioning cessation is imprecise, and the corresponding ambiguity as to the time that relaxation begins, we find no reliable sign of such a transition, even in samples of large grain size mortar similar to those described elsewhere as having clear and late cutoffs.

## I Introduction

Slow dynamic elasticity (SD) is a remarkable and universal non-classical nonlinear elastic behavior in which a modest mechanical conditioning induces a loss of stiffness that afterwards recovers, slowly, like log(time). This apparent healing after damage is seen in many materials, on length scales from the laboratory to the seismic, and on time scales from msec to years. Laboratory applications of minor strain (as little as $10^{-6}$) lead to immediate drops in elastic modulus that then slowly recover[1-9] over seconds to hours.  Loss of stiffness and slow recovery are seen also in seismic wave speed near a fault after an earthquake[10-12] where recoveries are monitored over periods from days to years. SD is observed in natural rocks, concrete and mortar[13,14] , and in buildings [15, 16]

Materials with simpler chemistry and structure show the effect as well. Cracked glass exhibits the behavior[17-19]. It is seen in unconsolidated aggregates of beads[20-23].  It is seen in an isolated bead confined between plates[24,25]. The inference is that unconsolidated materials replace slow dynamic processes at the internal inter-grain contacts of rocks with processes taking place at bead contacts.

There remains no consensus as to the mechanisms responsible.  It has been hypothesized [22] that there is connection to the better-studied phenomenon of log(t) aging of static friction [26] for which it is argued that contact areas and stiffness between grains will be dominated by asperities.  If the contact areas grow like log(t), as they do in certain models of plastic flow[26] and as has been observed [27]  in careful measurements, one derives log(t) aging of frictional strength, and presumably also elastic stiffness. It is also thought that moisture may play a role in SD, as it does in the linear elastic moduli of rocks.  Bittner[17] reported a strong humidity dependence of SD in cracked glass.  A role for moisture is further suggested by Bouquet *et al*'s observation[32] of log(t) aging and humidity dependence of the strength of a sand pile against avalanching.

A recurring hypothesis is the Arrhenius in which recovery proceeds[1,29,30] by way of bond-formation after thermally activated barrier penetration, as in [28] in which the bonds are water bridges, or to thermally activated plastic flow [28,29]. The Arrhenius hypothesis explains[30, 31,32]  log(time) aging by hypothesizing a distribution of barrier energies that is constant over a short range.  Logarithmic aging over many decades in time requires only constancy of that distribution over a short range in energy.  Amir et al[30] point out that a relaxation process that is the product of several subprocesses with random rates, will also tend to exhibit to log time behavior.

Until recently it has been understood that laboratory measured recoveries, when plotted vs the log of the time since conditioning cessation, are linear from several seconds to hours. But recent work[8, 33,34] has reported slopes that diminish at the earliest of times ( put differently, the relaxation spectrum is weaker for the fastest rates) and do so at different characteristic times for different materials. Kober et al [33] found that larger grain sizes were associated with more severe, and later roll-offs, i.e greater deficits in relaxation spectra, with reported spectral cutoffs at several seconds in some concretes. Gueguen et al[16] report slope diminishment in granite at times as late as several seconds.  These are important observations, as they challenge theorists and suggest potential application to material characterization and NDE, and perhaps point to mechanisms.  Figure 1 illustrates the deviation reported by Shokouhi et al [8].

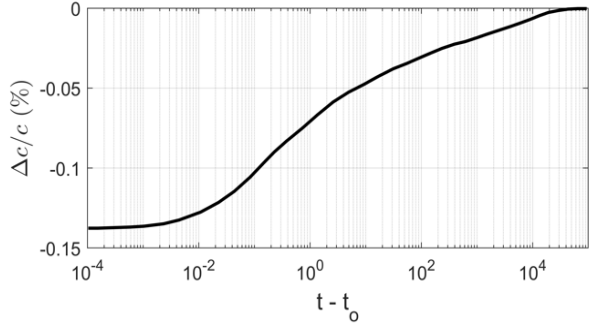

Fig 1] The data from Shokouhi et al's[8] fig 2 showing slow recovery of ultrasonic wavespeed c in Berea after vibration conditioning. The diminished slope at early times is evident. Their reference time $t_o$ was chosen 20 msec after the conditioning vibrations were interrupted, (15 msec after they appeared to have finished dying away.) That choice safely excluded all times before complete cessation of the ringdown of the conditioning but as argued here was excessively late, thereby leading to the apparent diminished slope.

That diminished slopes at early times have not been more widely observed has been attributed[8,33] to a dearth of experiments that can access sufficiently short times. The NRUS technique (nonlinear resonant ultrasound spectroscopy) in which changing stiffness is monitored by observing changing resonant frequency[1] does not lend itself well to tracking rapid changes or to fine time resolution; NRUS measurements have typically been confined to times greater than 10 sec.

Lobkis and Weaver[9] also reported deviations from log linearity at early times. Unlike those [8,33,16] cited above, they reported *increased* logarithmic slope at short times, discernable in cement paste for 3 < t < 50 msec and in sandstone for 3 < t < 300 msec.

Deviations from log(t) at timescales accessible in the lab may be perplexing for those who ascribe to the Arrhenius hypothesis. This hypothesis holds that the relaxation rates are given by an atomic scale attack rate of order $\nu=10^{12}$/sec, but diminished by the need to overcome high barrier energies E through rare thermal fluctuations. Assuming a distribution f(E) of barrier energies, one obtains, where k is Boltzmann's constant and T is temperature,

$$C(t) = C_\infty - \int f(E) \exp(-\nu t \exp(-E/kT)) \, dE \quad (1)$$

which implies an approximate local relation [32]

$$dC/d\ln t = kT < f(kT \{\ln \nu t + 0.577\}) > \quad (2)$$

Here C is any convenient measure of wavespeed or modulus. The brackets indicate a smoothing, or running average. Thus the hypothesis leads to log linearity when f is constant over a short range in energy. The hypothesis replaces the mystery of log linearity over decades in time (say from 1 to 1000 sec) with a lesser mystery of constant f over a short range in E (from 0.69 to 0.86 eV). By the same token, however, a slope dC/dlnt that significantly deviates from a constant as in fig 1 requires f(E) to vary significantly over small fractions of an eV. Such is not impossible, but arguably odd.

An equivalent expression [31] writes the recovering quantity C(t)

$$C(t) = C_\infty - \int A(\tau)/\tau \; \exp(-t/\tau) \, d\tau \quad (3)$$

in terms of a relaxation spectrum $A(\tau)$. Others [e.g. 33] define relaxation spectrum as $F(\tau) = A(\tau)/\tau$. Inasmuch as different relaxation times $\tau$ could be associated with different microstructural features or sizes, it could be worthwhile to retrieve the spectrum A. The inverse process of obtaining $A(\tau)$ from the recovery profile C(t) is, however, poorly posed, although it can be facilitated[8,33] by restricting the allowed forms for $A(\tau)$.

Deviations of C from log linearity at short times will correspond to features in A at low $\tau$. For this reason, such deviations[8,33,16], and their corresponding non trivial spectra [8,33], are deserving of replication. This report is intended to address that need.

The next section presents our measurements of SD recoveries at short times in a prism of Berea sandstone. Scrutiny of the data shows that ring-down of the conditioning pump vibrations can complicate the analysis, especially if the ring-down is prolonged. There is a consequent challenge to identify an effective start-time $t_o$ for the recovery. Regardless, however, of the ambiguity, we find that the data exclude deviations from log linearity for all times later than a few msec.

The provocative 2005 report by Lobkis and Weaver[9] of *increased* slopes at short times in sandstone and cement paste is then discussed and re-analyzed in section III. With allowance for ambiguity in $t_o$, we find no compelling evidence for decreased slope at short times.

Section IV presents a study of SD in the single aluminum bead, with focus on the early times. Again we find that slope diminishments can be excluded, in this case for all times greater than 20 msec.

Section V presents measurements of SD recoveries in a mortar prism with microstructures similar to one studied by Kober et al[33]. We find no slope diminishments consistent with those reported there.

**II Berea**

The test system for measurements on the Berea sandstone prism is shown in figs 2. It resembles that of the DAET (Dynamic Acousto-Elastic Testing] method [34, 8] but differs in two minor ways. The pump conditioning is applied by a Labworks' ET-126 electromagnetic shaker (rather than a piezoelectric disk) that is driven at a steady alternating current amplitude I, corresponding to force amplitudes αI with α = 4.5 Newtons/Ampere. It is driven longitudinally at fixed frequencies of order 2 to 12 kHz not necessarily on a structural resonance. Vibrations of the sample are monitored by an accelerometer on the tip.

The state of the material is probed using pulse/receive ultrasonics. Typically DAET [34] tracks the changing transit time, of order 6 μsec, of ballistic ultrasound across a sample. We instead monitor a (ca 500 kHz) diffuse wave that reverberates in the sample with a lifetime of order 500 μsec corresponding to path lengths of up to one meter. The transducers are placed approximately one quarter wavelength below the top where the vibration strain is greatest, but they do not face each other. The fractional dilation, or "stretch" of the received waveform is our measure of fractionally diminished wavespeed. Signal processing is described at greater length by Yoritomo and Weaver[23]. Ultrasonic pulses are launched with a 330 Hz repetition rate. Ultrasound and acceleration signals are acquired continuously at 10 MSa/sec for 100 seconds.

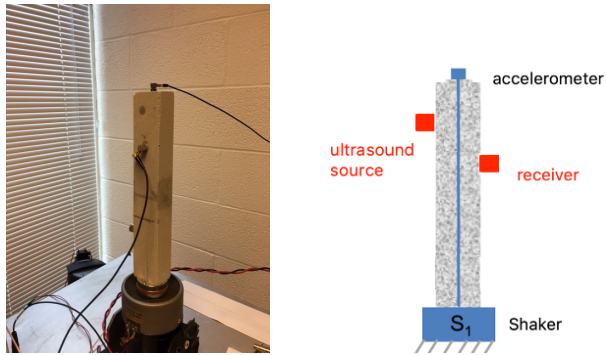

Fig 2. The laboratory system for studying SD in a prism of rock or mortar. The photo is of the Berea sample. The Berea sample is attached to the shaker head, and the transducers to the sample, through a removable glue.

Our block of Berea sandstone, grain size 125–350 $\mu$m, has dimensions a x b x L =35-37 x 47-50 x 305 mm. Its lowest free-free natural frequency is $f_1$ = 3.1 kHz and 2.95 kHz ( obtained respectively from a tap test and first resonance under harmonic forcing in the configuration of fig 2 ) From this it is calculated to have a bar-wave speed $(E/\rho)^{1/2} = c_{Bar} = 2Lf_1$ = 1830 m/s.

Figs 3 show the response of the Berea sample upon conditioning with harmonic vibrations at 2.5 Amperes (11.7 N) for 10 seconds at 3 kHz. Ultrasonic probe pulses were launched every 3msec, the signal from each resulting diffuse wave being used to calculate the stretch values (blue points). Fig (3a) shows the resulting stretch versus time over the 100 seconds of record. Fig (3b) shows both the stretch and the acceleration in the vicinity of conditioning cessation at t = 15.8 sec. Figs 3ab have several noteworthy features:

**1)** During the conditioning (between t =5 and 15 seconds in fig 3a) the diffuse waveform stretch, representing fractional change in ultrasonic wavespeed, drops quickly at first to about -0.005, and then continues to drop but more slowly. Continuing slow decreases are also seen elsewhere and indicate a slow loss of stiffness. The stretch furthermore shows something that looks like noise during the conditioning ( but is not - as can be seen by comparing with the low noise level for t < 5 sec, i,e., before the pump is turned on). The 'noise' is due to classical fast nonlinearity and has a variation that rarely exceeds ±0.00025, or ±5% of mean stretch -0.005. The magnitude of this 'noise' would be diminished if the probe waveform had a duration sufficient to average over one or more cycles of the pump, perhaps at a cost of decreased time resolution. Its noisy appearance is due to random phase differences between the pump waveform and the ultrasonic pulses.

**2)** The acceleration record (red curve in fig 3b) has a steady-state amplitude of $a_{ss}$ = 190 m/s$^2$ and exhibits no obvious sign of higher harmonics. We infer a strain amplitude $a_{ss}/c_{bar}\omega$ = 5.5μstrain at a point one quarter wavelength (equal here to 150 mm) below the tip. Higher harmonics appear at higher driving force (not shown). This is in contrast to other reports e.g.[8] whose plots do show such signs.

**3)** Fig 3b shows that the pump wave decays exponentially after the signal to the shaker is cut off, with a time constant of about 2msec.

**4)** Fig 3b shows three choices for the reference times $t_o$ used to construct $\log(t-t_o)$. The earliest is at the time at which the conditioning strain amplitude has dropped to 1/e of its steady state of190m/s$^2$. This occurs about 2 msec after the power to the shaker is cutoff, (blue dashed line). Another is near the 1/2e point of the exponential decay (red line about 2 msec later). A third (green line) is 4 msec later yet, after the pump vibrations have nearly died away.

**5)** There are two distinct and subtle issues related to the choice of $t_o$. One is to recognize that any measurements of instantaneous (albeit averaged over the duration of the probe wave) modulus, wave speed, or stretch can be contaminated by fast nonlinearities associated with simultaneous conditioning. The strength of this contamination to the stretches of interest can be estimated from how severely they fluctuated while conditioning was steady. The contamination can be mitigated by rejecting, or flagging as unreliable, any delays or stretches pertaining to times while ring down still has significant amplitude. One way to do this in a plot vs $\log(t-t_o)$, is to choose a $t_o$ well after the ring down - one that will naturally reject all data points where log t-

$t_o$ is undefined. This can have the unfortunate consequence of artificially generating a roll-off. But one need not use a late $t_o$ to enforce the rejections. It suffices to recognize and quantify any uncertainties in measured stretches. The uncertainty of a measure of stretch at a time t may be calculated from the observed variation while the conditioning was steady, times the ratio of the conditioning amplitude at time t to the amplitude while in the steady state; in this case, $U(t) = \pm 0.00025\, (\bar{a}(t)/a_{ss})$ proportional to the amplitude $\bar{a}$ of the acceleration at t. As the ringdown dies away, this uncertainty vanishes. For these data, the uncertainties are negligible except for the points at Stretch = -0.0025 and -0.0019, where they are ±0.00008 and ±0.00003 respectively.

**6**) A more complex issue is the unknown process by which, during ring-down, ongoing conditioning is competing with ongoing recovery. In plots vs log(t- $t_o$) we would like the plots to pertain to material properties, and so $t_o$ should be a time after the cessation of the conditioning and before the beginning of recovery. Yet during ring-down we surely have both simultaneously and there is no such $t_o$; the best one can do is argue for some range of plausible effective $t_o$. Absent a theory that purports to describe this domain of simultaneous damage and healing, we are at a loss. Nevertheless, we can examine a set of plausible effective $t_o$'s, and ask what deviations from log linearity would be consistent or inconsistent with that set.

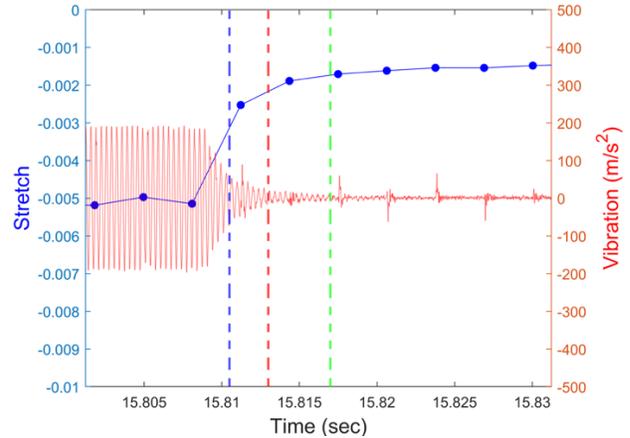

Fig 3b] A close-up of the stretches and the tip accleration near the time of vibration cessation in Berea. The periodic blips in the acceleration curve are due to electronic cross talk from the ultrasound.

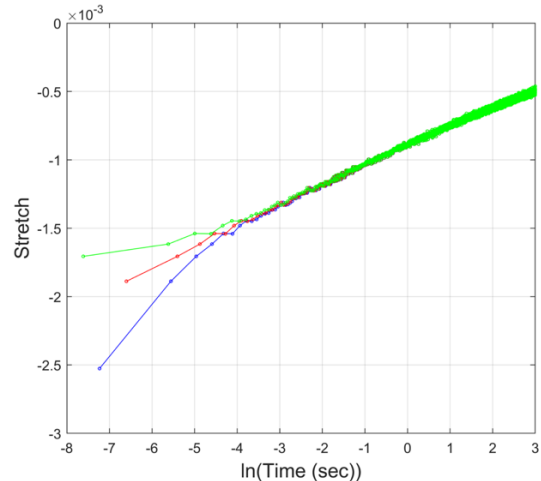

Fig 3c Plots of stretch versus ln(t-$t_o$) for the three choices of $t_o$ for Berea

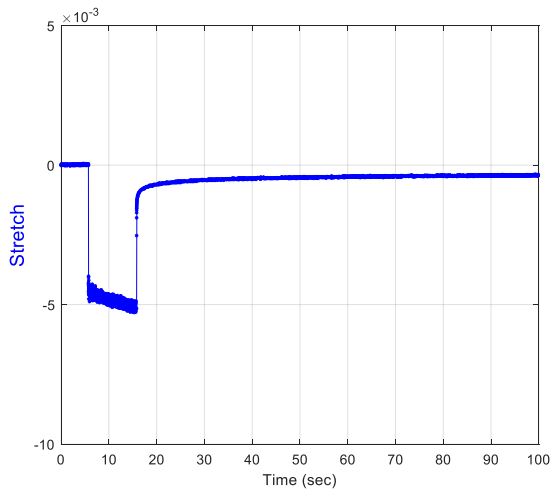

Fig 3a Plot of stretch S(t) in the Berea prism before during and after 10 sec of conditioning by 3 kHz vibrations with an peak amplitude of 5.5 μstrain.

Fig 3c plots the stretches versus ln(t-$t_o$) for the three choices of $t_o$.[35] The green (upper) line corresponds to the most delayed choice for $t_o$, at 15.817 sec. This choice, like ref[8]'s, saves the green curve from contamination by fast nonlinearity; none of its points suffer from significant uncertainties U. The choice does, however, lie well after the cessation of conditioning and the presumed begining of any relaxation, so the curvature in the green curve should not be taken as reflecting material properties. This is akin to Ref[8]'s choice of $t_o$ and consequent significant rolloff. To be sure, researchers may plot versus whatever they desire. The error, if any, will be in the interpretation. Curves like this are included here chiefly for illustration. We can nevertheless quantify any curve's report of a time scale for early-time slope diminishment by quoting the value of t-$t_o$ at which the slope is half what it is at later times. If we did this for the green curve it comes to about exp(-4.5) = 11 msec.

The red (middle) curve corresponds to the intermediate choice for $t_o$. Its first data point has stretch S = -0.00189, with an uncertainty ±0.00003. All of its other points have negligible uncertainty. The red curve runs though the middle of the error bars and shows no curvature. The point at which its slope is half that at later times is not identifiable but is safely judged to be before exp(-6.5) = 1.5 msec. Replacing the S values with the highest values consistent with the uncertainties, leads to a curve with slight upward curvature, and a time of half-slope at about exp(-6) = 2.5msec

The blue (lower) curve shows, if the uncertainties are neglected, *increased* slope at short times. If the S values are replaced with the highest values consistent with the uncertainties, one finds no sign of curvature in the blue curve, either up or down.

We conclude that no plausible choice for $t_o$ or adjustments in stretches S consistent with the uncertainties U can support times of half-slope greater than 2.5 msec. This is in contrast to the report[8] of a time of half slope equal to about 20 msec, but it is consistent with a private communication from Jan Kober of $\delta c(t)$ in Berea[33], in which, using a resonance tracking method for velocity changes $\delta c$ they found no rolloffs at times greater than 600 msec. (Earlier than 600 msec they had no data, as that probe method was ill suited to investigating shorter times.) They also used a different method to probe stiffnesss changes, and converted those to an effective relaxation spectrum $F(\tau)$ by inverting Eq 3. We can reconstruct their $\delta c$ data by integrating Eq 3 using the $F(\tau)$ data from their fig 4; this gives a time of half-slope (half the maximum slope) at about 50 msec, far greater than the 2.5 msec bound reported here.

### III  SD in cement paste and sandstone after impact

The earliest work to report slow dynamics at short times was that of Lobkis and Weaver[9] for blocks of cement paste and sandstone, with examined times from 3 msec to 400 sec, over 5 decades in time. Conditioning was provided by the impact of a wooden ball bearing; the acoustic emission from which provided the time of impact accurate to within microsecs. Elastic stiffness changes were assessed by monitoring a 525 kHz Larsen frequency (the steady state harmonic screech provided by an ultrasonic version of the familiar audio screech heard when microphones and speakers are too close and/or gain is too high.) The SD recoveries were found to be linear in log(t) for late enough times. Signs of deviation (with *increased* slope) from linearity in log( t-$t_{impact}$ ) were apparent for times at and before 50 msec in cement paste, and 135msec in sandstone. See the lower curves of figs 4a and 4b.

To our knowledge this is the only report of accelerated recovery at short times, and is in apparent direct conflict with reports elsewhere[8,16,33] of slower recovery at short times. For this reason further inquiry is especially interesting. Was the observation[9] meaningful?

It is not clear that the time of impact $t_{impact}$ corresponds to the end of significant conditioning and beginning of recovery. The sample surely vibrated for some time after the impact. The end of conditioning is therefore at some time at or after the impact. We can, though, replot that paper's data using various guesses for $t_o$. Figs 4a and 4b show the result of that operation.

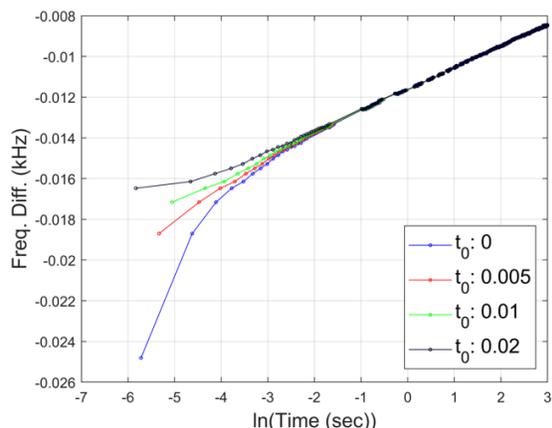

Fig 4a  A plot of the data of ref[9]'s fig 11a showing the evolution of a Larsen frequency after the impact of a 4mm wooden ball on cement paste. Different choices for $t_o$ (in seconds after ball impact) result in different curves. We remark that the lower, blue, curve's first point corresponds to a time only 3 msec after the impact, and is arguably contaminated by classical fast nonlinearities from the impact's acoustic emission ring down ( see fig 5 ). One could also argue that the continuous Larsen wave time-averages over the oscillations impact and is therefore not sensitive to fast nonlinearities

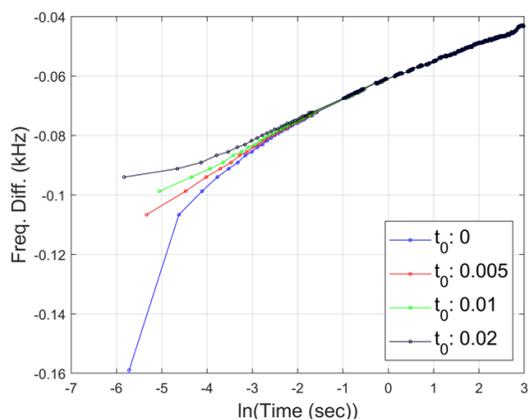

Fig 4b] A re-plot of the data of Lobkis and Weaver's[9] fig 11b showing the evolution of a Larsen frequency after the impact of a 4mm wooden ball on a sample of sandstone (type unknown). Different choices for $t_o$ (in seconds after ball impact) result in different shapes.

Fig 4a shows four curves corresponding to four choices for $t_o$. The bottom (blue) curve is that plotted by Lobkis and Weaver[9], with reference time taken at the instant of ball impact. The increased slope at short times is evident. The lower-middle (red) curve is for a reference time $t_o$ 5 msec later; the green (upper middle) for 10 msec after the impact; the black (top) for 20 msec after the impact. The data indicate that, if one is willing to entertain a time 5 or 10 msec after impact as the effective moment at which conditioning ceases and recovery begins, then there is no sign of deviation from log linearity. One must choose an even later $t_o$ if short time diminished slope is to be asserted.

Fig 4b is like 4a except that it corresponds to Lobkis and Weaver's[9] sandstone sample. The conclusions are unaltered. In particular, a choice of 5 or 10 msec again serves to remove early time deviations from log linearity.

We would like to know which of these $t_o$, if any, best describes an effective end of conditioning and beginning of recovery. To inform that judgement, the wooden ball was dropped on the original cement sample, and the resulting vibrations detected with an accelerometer. That signal was high pass filtered at 600 Hz and integrated to give material velocity. See fig 5. The envelope rises sharply at the impact and then decays, initially with a time constant of order 2msec. But a low frequency wave persists for many tens of msec. Conditioning could in principle continue beyond the time of impact, due to the stresses during the vibrations. It is difficult to judge the relative strength of conditioning due to the impact (with large strain but confined to short duration and extent, estimated[9] as of order 1% and 300 μm respectively) and the longer lasting lesser strain (well below 1 μstrain) in the more extended reverberant vibrations. Setting $t_o$ to 5 msec after impact is perhaps reasonable. Setting it at 20 msec, necessary for a the top curves' diminished slopes, would demand admitting the long lived very low amplitude as a significant agent of conditioning. We conclude that there is no evidence for a short time diminishment, and some uncertain evidence for a steepening.

Perhaps the chief lesson to be taken from this is that impact conditioning is problematic; it can generate low frequency reverberant strains whose slow decay obscures identification of reference time $t_o$. It is also noteworthy that impact pump strains of order 1%[9] are far greater than are typical in the SD literature, and that the unusual steepening at short time may be related.

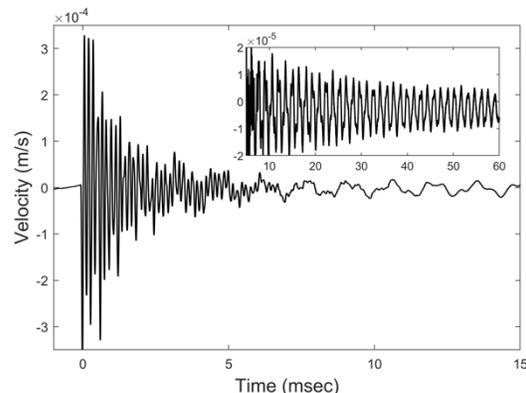

Fig 5] Accelerometer response (after high pass filtering and integration) to the impact of a 4mm diameter wooden ball on the cement bar of Lobkis and Weaver[9]. Signal decays quickly over the first few msec, but a low amplitude 700 to 1700 Hz component persists for much longer. The strains associated with these material velocities are estimated using $\varepsilon = v/c_{bar}$ [36] to be very small, of order 10 to 100 nanostrain.

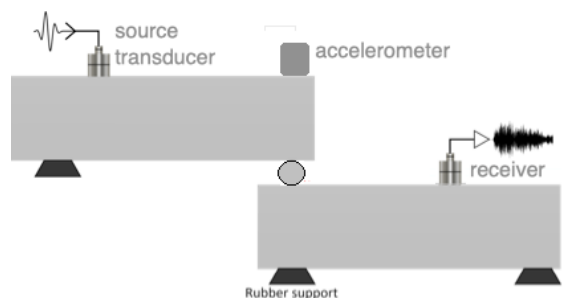

Figure 6] The single bead experiment (not to scale).

## IV Early time in single beads

Figure 6 shows the system for measuring SD in a single bead. The structure sits on vibration isolation and optical tables to minimize the influence of ambient building vibrations. An ET-126 electromagnetic shaker also sits on that table and provides 500 Hz conditioning vibrations, typically for 30 seconds. An accelerometer placed on the upper slab over the bead, reports amplitudes of order $1m/s^2$ while the shaker is on and provides a proxy for the dynamic conditioning force on the bead. The bead/slab contact stiffness is diminished by the vibrations, but then recovers after the conditioning ceases. Ultrasonic pulses are introduced to the upper slab at rates up to 20 Hz. Any faster and the diffuse wave from one pulse can overlap that from the next. The resulting diffuse field leaks into the lower slab via the bead, whose transmissibility is dominated by the stiffness of its contacts. When that stiffness diminishes, the diffuse signal in the lower slab is delayed. Coda wave interferometric methods extract that delay.

This system, with pulse repetition rates of order 1 Hz, was first introduced by Yoritomo and Weaver [24,25]. Weaver and Lee[37] discussed the theory relating measurements of delay to stiffness changes. They also[32] investigated the effect of dynamically heating the bead. Throughout that work, log(t) recoveries were routinely observed, and with low noise. But their slow pulse repetition rate, now improved, prevented investigation of the behavior at the shortest times.

Fig 7a plots the diffuse wave delays in a system consisting of an aluminum bead and slabs. 90 seconds of reference data is followed by 20 seconds of conditioning. Average delays during conditioning are obscured by rapidly varying large positive and negative 'noise,' due to fast nonlinear dynamics similar to that in fig 3a but stronger here. Vibrations cease at t =110s, after which the bead is left with a diminished stiffness. The stiffness then recovers, quickly at first and then increasingly slowly. Time stamps for the delays Y have been set at the midpoint of the 10 msec segment of the received diffuse waveforms that generate Y. Fig 7b presents the same data on an expanded scale in the vicinity of the vibration cessation time. It also shows the accelerometer signal whose 500 Hz vibrations (red curve) do not cease instantaneously, but ring down with a time constant of about 15 msec. Three choices for reference time $t_o$ are indicated by the vertical dashed lines. Fig 7c plots the delays vs log($t-t_o$) for the three choices of $t_o$. The red (upper) curve is for the most delayed of the three choices, $t_o$ = 140 msec after the beginning of the ring-down. It shows a diminishment of slope at short times reminiscent of that reported elsewhere[8]. But its $t_o$ is manifestly long after the conditioning has ended, so its slope diminishment should not be considered material but rather an artifact of the late choice for $t_o$. The green (middle) curve shows good linearity down to its limit at 25 msec. The blue curve is based on another plausible choice for $t_o$ and shows a slightly *increased* slope at short times. The net conclusion is that these data do not support diminishments in slope, at least at times later than 25 msec.

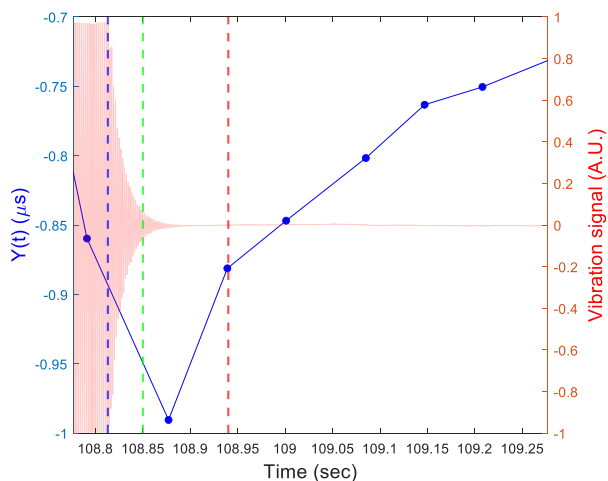

Fig 7b] A close examination of the data of figure7a in the vicinity of the time of vibration cessation.

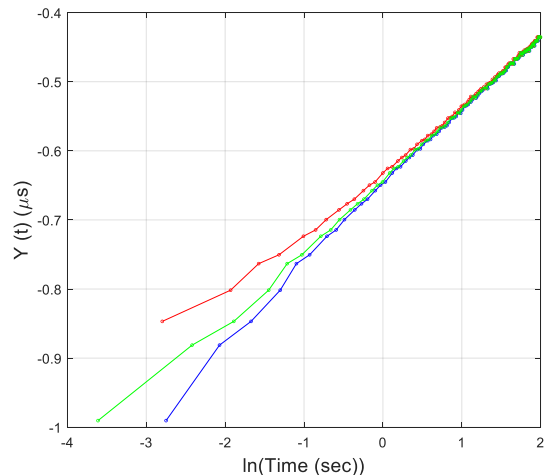

Fig 7c] The single bead data of fig 7a is plotted vs log($t-t_o$) for the three suggested values of $t_o$.

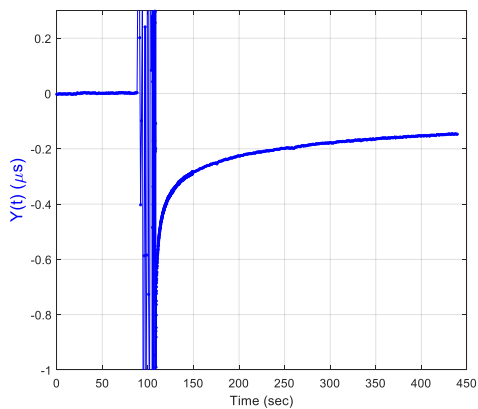

Fig 7a The shifts Y in a single bead aluminum system vs time before during and after 10 sec of conditioning. Negative Y represents delay.

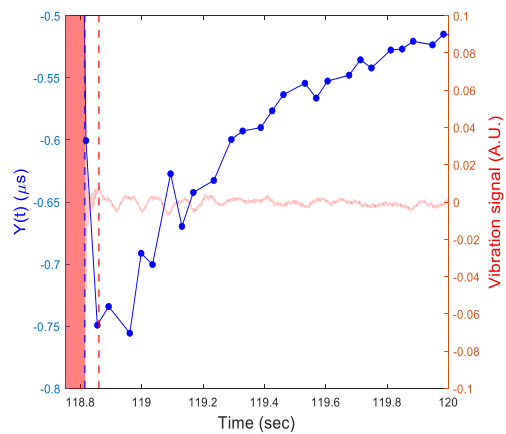

Fig 7d. A single bead test in glass. Residual low frequency vibrations contaminate the delays Y with fast nonlinearity.
7

Fig 7 d shows the data from a similar test in single glass bead and slabs. Residual ~10 Hz vibrations of unknown origin noticeably contaminate the measured delays with fast nonlinearities. Positive and negative excursions of the delays Y correlate with negative and positive values of the acceleration signal at late times.

**V Slow dynamics in a mortar prism.**

A mortar prism of dimensions 25.4 x 25.4 x 294 mm was constructed consisting of water and Portland cement and sand in a ratio by weight of 0.5 : 1 : 3. Sand was sieved to a grain size range of 2.36~4.75 mm. The sample was cured for two weeks at elevated humidity in order to minimize cracking, and aged for months. Such a sample is of particular interest here, as Kober *et al* reported the strongest deviations from log linearity in this kind of material.

The experimental system is like that used for the Berea and illustrated in fig 2. The sample was epoxied to an aluminum disk that was then bolted to the shaker head.

The sample was driven by 3.8 Amperes (16 Newtons) at 3.8 kHz near the first resonance. Figs (8) show results from the mortar sample. Acceleration amplitude was about 900 m/s$^2$. The absolute noise level is about the same as in Berea (figs3). S/N is lower, which is attributed to the mortar having weaker slow dynamics while the noise in stretch is unchanged. The acceleration signal (red curve in 8b) shows the bar to be in a periodic but not harmonic steady state, due to a classical fast nonlinearity not present in fig 3b.

The points in fig 8b at Stretch = -1.7 and -2.7 x10$^{-3}$ are likely well contaminated by the dying away vibrations. The large fluctuations in stretch during steady state conditioning between 5 and 10 sec in fig 8a, notwithstanding the much lower conditioning amplitude at these later times, imply that these data points are too uncertain to be included in further analysis. They are therefore excluded from fig 8c.

As in previous sections, we find that the choice of $t_o$ affects the curves. Nevertheless, a claim of early time slope diminishment can only be sustained if one chooses the unlikely looking latest $t_o$ (green curve). Even then the resulting half-slope point occurs only for $t-t_o < \exp(-6) = 2.5$msec.

This bound of 2.5 msec is far far earlier than the corresponding time for that point (at ~4 seconds) in the Kober et al [33] sample Conc-B06 grain size 10 mm (see their fig 1) at which the slope dC/dlnt goes to its half amplitude.

2.5 msec is also far far earlier than the corresponding time[33] for the half-slope point (at about 2 sec) for their sample Conc-X04 with grain size 5 mm. We constructed this estimate by performing the integration, Eq 3, on the data F($\tau$) from their fig 4's F($\tau$) in their fig 4 of A/t for their sample.

This test was repeated after a heat treatment to the sample (90°C for 12 hours). The heat treatment led to a slower ring down, and though the conditioning strain amplitude was half as much, the SD slope was doubled. Conclusions about recovery roll-offs were unaltered.

While there are differences in material and samples and measurement methods, both in pumping and probing, the three-order-of-magnitude discrepancies seem larger than can be explained by such considerations. Nevertheless we list some of those differences:

It is apparent that the roll-offs in [8] may be ascribed to its analysis's excessively late choice for reference time. But that will not explain the discrepancy with Kober et al [33] who choose $t_o$ at the moment of the electronic cutoff to the conditioning. Were we to have made such choice, all our curves, i.e, like the upper curves in figs 3c, 4ab, 7c and 8c, would curve *up*, with *increased* slope at earliest times.

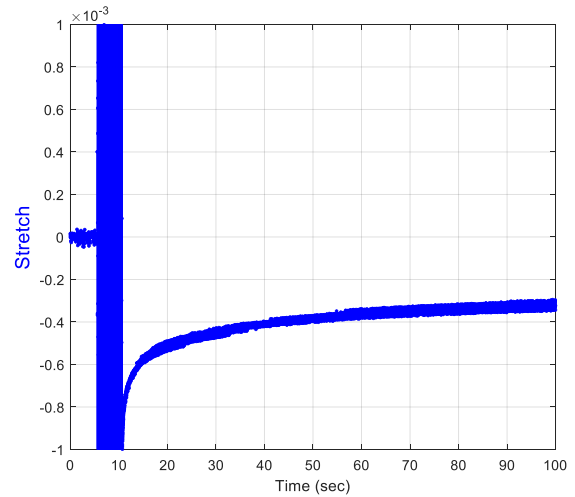

Fig 8a Stretches before during and after 5 seconds of 3.8kHz 3.8 Ampere conditioning on a sample of 2.36~4.75 mm grain size mortar.

Kober et al condition their concrete samples with longitudinal vibrations at tens of kHz in the vicinity of a resonance, for 5 minutes. They then probe, using their protocol I, with a low amplitude continuous sinusoid at the same frequency, and monitor the phase and amplitude of the sinusoidal response to extract the wave speed. This is a complex process and very different from ours. We condition for 5 seconds and probe with pulsed ultrasound, extracting wave speed changes by means of coda-wave interferometry.

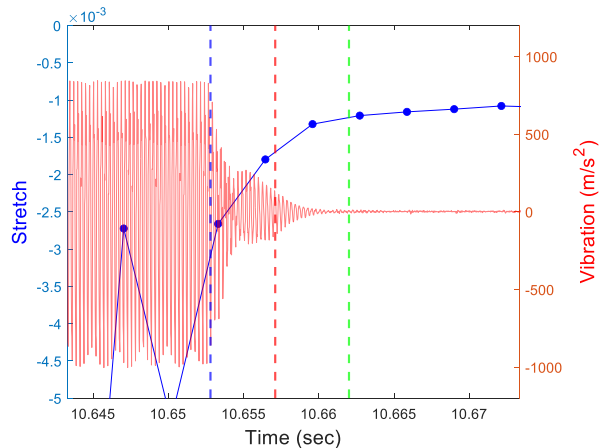

Fig 8b. A close up of the data of fig 8a near the time of vibration cessation, together with the bar's tip acceleration. Three choices for $t_o$ are indicated.

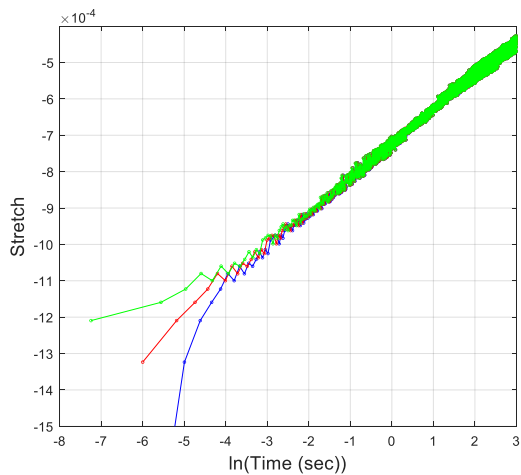

Fig 8c] The stretch data for the mortar prism of fig 8a, versus $\ln(t-t_o)$ for the three choices of $t_o$.

## VI Conclusions

We have examined the earliest times of SD recovery in a variety of systems. This includes concrete and Berea systems for which significant early time roll-offs have been reported by others. While a degree of uncertainty in the proper choice for reference time $t_o$ needed to construct $\ln(t-t_o)$ is inescapable, we find that no plausible choice thereof leads to roll-offs remotely comparable to those reported elsewhere. The data are in fact consistent with no early time deviations from log linearity in any of our materials. They further indicate that if there are diminishments in recovery slope at short times, they must occur at times earlier than 2.5 msec.

We offer no explanation for the differences between these measurements and others', noting only that samples and materials are never fully identical, nor the methods by which those samples are pumped and probed, nor the choices of $t_o$, nor the durations of the conditioning ringdowns. It would prove useful to determine the source of the differences.

## Acknowledgements

The work was supported by the U.S. Department of Energy, Office of Science, Office of Basic Energy Sciences, under Award No. DE-SC0021056. The authors are grateful to John Popovics for providing samples, and to Jan Kober for providing the data for his figs 1 and 4.